\documentclass[11pt]{article}
\usepackage{latexsym,amssymb,amstext,amsmath}
\usepackage{slashed}
\usepackage{mathrsfs}
\usepackage{color}
\usepackage{hyperref}

\allowdisplaybreaks
\newcommand{\ft}[2]{{\textstyle\frac{#1}{#2}}}
\newcommand{\bea}{\setlength\arraycolsep{2pt} \begin{eqnarray}}
\newcommand{\eea}{\end{eqnarray}}
\newcommand{\nn}{\nonumber}

\usepackage{hyperref}

\def\rmi{{\rm i}}
\newsavebox{\uuunit}
\sbox{\uuunit}
{\setlength{\unitlength}{0.825em}
	\begin{picture}(0.6,0.7)
	\thinlines
	\put(0,0){\line(1,0){0.5}}
	\put(0.15,0){\line(0,1){0.7}}
	\put(0.35,0){\line(0,1){0.8}}
	\multiput(0.3,0.8)(-0.04,-0.02){12}{\rule{0.5pt}{0.5pt}}
	\end {picture}}

\newcommand{\U}{\mathop{\rm U}}
\newcommand{\SU}{\mathop{\rm SU}}

\def\be{\begin{equation}}
\def\ee{\end{equation}}
\def\ba{\begin{array}}
\def\ea{\end{array}}
\def\bea{\begin{eqnarray}}
\def\eea{\end{eqnarray}}
\def\bd{\begin{displaymath}}
\def\ed{\end{displaymath}}

\def\nn{\nonumber}

\def\a{\alpha}

\def\g{\gamma}

\def\d{\delta}

\def\e{\epsilon}
\def\ve{\varepsilon}
\def\f{\phi}
\def\vf{\varphi}

\def\p{\psi}

\def\l{\lambda}
\def\L{\Lambda}
\def\m{\mu}
\def\n{\nu}
\def\r{\rho}
\def\s{\sigma}

\def\o{\omega}

\def\nn{\nonumber}
\def\cD{\mathcal{D}}
\def\cN{\mathcal{N}}
\def\cC{\mathcal{C}}
\def\cL{\mathcal{L}}
\def\cF{\mathcal{F}}

\makeatletter
\@addtoreset{equation}{section}
\makeatother

\begin{document}
%%%%%%%%%%%%%%%%%%%%%%%%%%%%%%%%%%%%%%%%%%%%%%%%%%%%%%%%%%%%%%%%%%%
%
\begin{titlepage}
%%%%%%%%%%%%%%%%%%%%%%%%%%%%%%%%%%
\begin{flushright} \small
  ITU-2016-002
\end{flushright}
%%%%%%%%%%%%%%%%%%%%%%%%%%%%%%%%%%
\bigskip

\begin{center}
  {\LARGE\bfseries Off-Shell $\cN=2$ Linear Multiplets in Five Dimensions}
  \\[10mm]
\textbf{Mehmet Ozkan}\\[5mm]
\vskip 6mm
{\em  Department of Physics, Istanbul Technical University,  \\ 
	Maslak 34469 Istanbul, Turkey, }\\[3mm]
 {\tt ozkanmehm@itu.edu.tr} 
\end{center}

\vspace{3ex}

\begin{center}
{\bfseries Abstract}
\end{center}
\begin{quotation} \noindent
We present a superconformal tensor calculus for an arbitrary number of five dimensional $\cN=2$ linear multiplets. We also demonstrate how to construct higher derivative invariants, and higher order supersymmetric off-diagonal models. Finally, we show the procedure required for the derivation of the supersymmetric completion of the non-Abelian $F^4$ action.

\end{quotation}

\vfill

%%%%%%%
\flushleft{\today}
%%%%%%
\end{titlepage}
\setcounter{page}{1}
\tableofcontents

%\newpage

%%%%%%%%%%%%%%%%%%%%%%%%%%%%%%%%%%%%%%%%%%%%%%%%%%%%%%%%%%%%%%%%%%%
\section{Introduction}{\label{Intro}}
%%%%%%%%%%%%%%%%%%%%%%%%%%%%%%%%%%%%%%%%%%%%%%%%%%%%%%%%%%%%%%%%%%%

The construction of supergravity theories can be realized either on-shell or off-shell. When on-shell, the construction procedure can be done by the iterative Noether procedure. In this case, the closure of the superalgeabra on the components of the supergravity multiplet requires the use of the equations of motion. At the two derivative level, and when matter couplings are ignored, the Noether procedure can be the simplest and the most straightforward way to obtain a supergravity model. When higher derivative couplings, or matter couplings are taken into account, the off-shell formulation provides a novel framework. In this case, the transformation rules of the supergravity and the matter multiplets are independent of the field equations, but the price to pay is the necessity to introduce auxiliary fields to match the bosonic and fermionic degrees of freedom, and for the off-shell closure of a multiplet. As an off-shell construction method, the superconformal tensor calculus \cite{Kaku:1978nz,Kaku:1978ea,Ferrara:1977ij,Kaku:1977pa},  which is based on an extension of the super-Poincar\'e algebra to include the superconformal generators, greatly simplifies the construction procedure. In this setting, one first constructs actions that are invariant under the superconformal symmetries, and then gauge fixes the conformal symmetries to obtain a Poincar\'e theory. 

In the case of five dimensions, the conformal supergravity is based on the exceptional superalgebra $F^2 (4)$ \cite{Nahm:1977tg}. A multiplet that contains the gauge fields of this algebra is called a Weyl multiplet. In five dimensions, there are two possible formulations of Weyl multiplets: the standard and the dilaton Weyl multiplets \cite{Kugo:2000hn,Bergshoeff:2001hc,Fujita:2001kv}. These multiplets have the same gauge fields, but they differ in the matter fields that one needs to introduce to match the bosonic and the fermionic degrees of freedom. In addition to the Weyl, the multiplets that are relevant to the superconformal construction procedure are the vector, the hyper and the linear multiplets \cite{Kugo:2000hn,Fujita:2001kv,Bergshoeff:2002qk}. When the standard Weyl multiplet is utilized, both the vector and the linear (or hyper) multiplets have to be used as compensators to gauge fix the redundant conformal symmetries. The reason for that is because the standard Weyl multiplet does not contain a graviphoton which makes the use of vector multiplet necessary, and the vector multiplet action by itself does not describe a consistent supergravity theory, which can be overcome by the use of either a hypermultiplet \cite{Kugo:2000hn,Fujita:2001kv,Bergshoeff:2002qk,Bergshoeff:2004kh,Kugo:2000af} or a linear multiplet \cite{Ozkan:2013nwa} as a second compensator. (A similar phenomena also occurs in six dimensional $\cN = (1,0)$, see \cite{Bergshoeff:1985mz,Coomans:2011ih}). When the dilaton Weyl multiplet is used, the multiplet itself contains a graviphoton, therefore the use of a vector multiplet is not necessary. Consequently, a single linear multiplet is sufficient to describe a conformal supergravity \cite{Coomans:2012cf}. 

As mentioned, the off-shell formulation is also of great use when the higher derivative extensions are considered. For a standard Weyl multiplet, the Weyl squared invariant \cite{Hanaki:2006pj} can be constructed with a vector multiplet compensator, however the Ricci scalar squared invariant \cite{Ozkan:2013nwa} requires a linear multiplet compensator. For the dilaton Weyl multiplet, the supersymmetric completion of the Gauss-Bonnet combination \cite{Ozkan:2013uk} and the Riemann squared action \cite{Bergshoeff:2011xn} do not require a compensator multiplet, however the Ricci scalar squared requires a linear multiplet compensator. Furthermore, if one is after a five dimensional gauged $\cN=2$ supergravity from an off-shell viewpoint, then the linear multiplet plays a crucial role to define a third Weyl multiplet: the deformed dilaton Weyl multiplet \cite{Coomans:2012cf} (see \cite{Butter:2014xxa} for the superspace construction of the deformed dilaton Weyl multiplet). 

Although the linear multiplet is an important ingredient of the superconformal gravity in five dimensions, there is no detailed investigation on the general couplings of this multiplet. Our aim in this paper is to fill this gap following \cite{deWit:2006gn}. As a byproduct, we also present the higher derivative off-diagonal invariants of five dimensional $\cN=2$ supergravity: the supersymmetric invariants that do not contain pure curvature tensor terms, but contain a curvature tensor multiplied by an auxiliary scalar. These invariants play an important role in obtaining ghost-free models of supergravity when the maximally symmetric vacua is given by $AdS$. In such cases, the elimination of the auxiliary field spoils the ghost-free combination of cuvature invariants as the equation of motion for the auxiliary field contains curvature terms sourced by the off-diagonal terms, see \cite{Bergshoeff:2010mf,Bergshoeff:2014ida} for three dimensional $\cN = 1$ examples. Therefore, it is necessary to construct off-diagonal invariants to obtain ghost-free supergravity models in an $AdS$ background. In the case of five dimensions, the ghost free higher derivative model is the Gauss-Bonnet combination \cite{Ozkan:2013uk}, and the necessary off-diagonal invariant is the $RN$ invariant where $R$ is the Ricci scalar and $N$ is the auxiliary field.

This paper is organized as follows. In Section \ref{Section1}, we review the rigid superconformal vector multiplet and construct and action for $N-$number of interacting linear multiplets of $\cN=2$ supersymmetry. In Section \ref{Section2}, we generalize the results of rigid conformal model to conformal supergravity couplings. In Section \ref{Section3}, we discuss the construction of a Poincar\'e invariant supergravity theory, and the off-diagonal invariants by use of linear and vector multiplets. In this section, we also comment on the supersymmetric completion of the $F^4$ action. We present our conclusions in Section \ref{Conclusions}.

%%%%%%%%%%%%%%%%%%%%%%%%%%%%%%%%%%%%%%%%%%%%%%%%%%%%%%%%%%%%%%%%%%%
\section{Rigid Superconformal Linear Multiplets}{\label{Section1}}
%%%%%%%%%%%%%%%%%%%%%%%%%%%%%%%%%%%%%%%%%%%%%%%%%%%%%%%%%%%%%%%%%%%

The five dimensional $\cN=2$ linear multiplets can be realized superconformally in a flat Minkowski background. In its bosonic sector, a linear multiplet  consists of $\SU(2)$ triplets $L_{ij}$, a constrained vector $E_a$, and an auxiliary scalar $N$. The fermionic  sector consists of an $\SU(2)$ doublet field $\vf_i$. The $Q-$supersymmetry $(\e^i)$, $S-$supersymmetry $(\eta^i)$, dilatations $(\L_D)$ and $\SU(2)$ R-symmetry $(\L^{ij})$ transformation rules for a linear multiplet  are given by\footnote{In this paper, we use the conventions of \cite{Ozkan:2013uk,Freedman:2012zz}}  \cite{Bergshoeff:2002qk}
\bea
\d L^{ij} &=& \rmi \bar{\e}^{(i} \vf^{j)} + 3 \L_D L^{ij} - 2 \L^{(i}{}_k L^{j)k} \,,\nn\\
\d \vf^{i} &=&  -\ft12 \rmi \slashed{\partial} L^{ij} \e_j - \ft12 \rmi \g^a E_a  \e^i + \ft12 N \e^i + 3 L^{ij}\eta_j + \ft72 \L_D \vf^i - \L^i{}_j \vf^j  \,,\nn\\
\d E_a &=&  -\ft12 \rmi \bar{\e} \g_{ab} \partial^b \vf  - 2 \bar{\eta} \g_a \vf  + 4 \L_D E_a   \,,\nn\\
\d N &=& \ft12 \bar{\e} \slashed{\partial} \vf  + \ft32 \rmi \bar{\eta} \vf  + 4 \L_D N     \,,
\label{RigidConformalLinear}
\eea
where the closure of the superconformal algebra on the components of the linear multiplet requires $\partial^a E_a  = 0$. Therefore, $E_a$ can be solved in terms of a 3-form $E_{abc}$ as 
\bea
E^a = - \ft1{12} \e^{abcde} \partial_b E_{cde} \,.
\label{ERigidSolution}
\eea
One can also define a dual 2-form potential $E_{ab}$ as $E_a = \partial^b E_{ab}$ and $E_{abc} = \e_{abcde} E^{de}$. The components of the linear multiplet are inert under the special conformal transformations $(\L_{K\m})$.

In principle, with the transformation rules (\ref{RigidConformalLinear}) in hand, one can start from a conformally invariant kinetic term for $L_{ij}$ and apply the Noether procedure to construct a two-derivative rigid superconformal action for linear multiplets. This method is very tedious, and there is a rather simpler way which requires the use of another conformal multiplet of $\cN=2$ supersymmety; the vector multiplet. The vector multiplet consists of a scalar field $\r$, a vector field $A_\m$, and $\SU$(2) triplet of auxiliary fields $Y_{ij}$. The fermionic sector consists of an $\SU(2)$ doublet field $\l_i$. The vector multiplet is inert under the special conformal transformations, and the $Q,S,D$ and $\SU(2)$  transformation rules are given by \cite{Bergshoeff:2002qk}
\bea
\d\r &=& \ft12 \rmi \bar{\e} \l  + \L_D \r    \,,\nn\\
\d A_\m &=&  \ft12 \bar{\e} \g_\m \l  \,,\nn\\
\d \l^{i} &=& -\ft14 \g \cdot F \e^i - \ft12 \slashed{\partial} \r \, \e^i - Y^{ij} \e_j + \r \eta^i  + \ft32 \L_D \l^i - \L^i{}_j \l^j    \,,\nn\\
\d Y^{ij} &=& \-\ft12 \bar{\e}^{(i} \slashed{\partial} \l^{j)}  + \ft12 \bar{\eta}^{(i} \l^{j)} + 2 \L_D Y^{ij} - 2 \L^{(i}{}_k Y^{j)k}  \,,
\label{RigidConformalVector}
\eea
where the field strength $F_{\m\n}$ is defined as $F_{\m\n} = 2\partial_{[\m} A_{\n]}$. The rigid conformal actions in five dimensions can be constructed based on the observation that a vector and a linear multiplet can couple linearly \cite{Kugo:2000hn}
\bea
\cL &=& Y^{ij} L_{ij} + \rmi \bar\l \vf + A^a E_a + \r N \,,
\label{RigidDensityFormulae}
\eea
where the gauge invariance of the action is satisfied with the condition $\partial^a E_a = 0$. As it stands, this Lagrangian, which we shall refer to as the rigid density formulae, does not mean much since the field equation derived from this action states that all the components of the vector and the linear multiplets are zero. However, we can regard the vector multiplet components not as fundamental fields but as composite expressions given in terms of the elements of the linear multiplet that transforms exactly as a vector multiplet (\ref{RigidConformalVector}).

For the construction of an action for the rigid conformal linear multiplets, let's define a real function $\cF_{AB} (L)$ of the scalars of the linear multiplets $L_{ij}^A$, where $A,B = 1,2,\ldots, N$ indicates the number of linear multiplets. The lowest component of the vector multiplet $\r_A$ can then be expressed in terms of the components of the linear multiplets as
\bea
\r_A &=& 2 \cF_{AB} N^B - \rmi \cF_{ABC}{}^{ij} \bar{\vf}_i^B \vf_j^C \,,
\label{LowestLinear}
\eea
where we have the following definitions
\bea
\cF_{ABC}{}^{ij} &=& \frac{\partial \cF_{AB}}{\partial L_{ij}^C} \,, \qquad \cF_{ABCD}{}^{ijkm} = \frac{\partial^2 \cF_{AB}}{\partial L_{ij}^C \partial L_{km}^D} \,.
\eea
The composite expression should transform exactly as the scalar component of the vector multiplet. The dilatation transformation of $\r$ implies that $\cF_{AB}$ must be of scaling dimension $-3$
\bea
\d_D \cF_{AB} = - 3 \L_D \cF_{AB} \,.
\eea
Furthermore, $S$-invariance of the $\r_A$ implies that
\bea
\cF_{ABC}{}^{ij} &=& \cF_{A(BC)}{}^{ij} \,, \qquad \cF_{ABC}{}^{ij} L_{jk}^C = - \ft12 \d^i{}_k \cF_{AB} \,.
\eea
Note that there is no particular symmetry in the indices of $\cF_{AB}$, however its derivative, $\cF_{ABC}$, must be symmetric in the second and the third indices. These constraints and the lowest composite expression (\ref{LowestLinear}) are consistent with the single multiplet construction given in \cite{Coomans:2012cf}   which corresponds to the choice $\cF_{AB} = \d_{AB} L^{-1}$ where $L^2 = L_{ij} L^{ij}$. Other examples that corresponds to other special choices of the generic $\cF_{AB}$ can be found in \cite{Butter:2014xxa}. The $Q$-transformation of the composite expression for $\r_A$ leads to the composite expression for $\l_{iA}$
\bea
\l_{i A} &=& - 2\rmi \cF_{AB} \slashed\partial \vf_i^B + 2 \cF_{ABC ij} \vf^{jB} N^C + 2 \rmi \cF_{ABC ij} \slashed{E}^B \vf^{j C} \nn\\
&& - 2 \rmi \cF_{ABC}{}^{jk} \slashed\partial L_{ij}^B \vf_k^C  - 2 \rmi \cF_{ABCDij}{}^{kl} \vf^{j D} \bar{\vf}_k^B \vf_l^C \,.
\eea
The last condition on $\cF_{AB}$ is related to the higher order spinor terms,
\bea
\ve_{jk} \cF_{ABCD}{}^{ijkl} = 0 \,,
\label{HigherOrderCondition}
\eea
which is needed to set $\cF_{ABCD}{}^{ijkl} \vf_k^D \bar{\vf}_i^C \vf_j^B = 0$ in order to ensure that the composite expression for $\l_{Ai}$ indeed describes the fermionic sector of a vector multiplet. More discussions on the function $\cF_{AB}$ in the context of four dimensional $\cN=2$ supersymmetry can be found in \cite{deWit:2006gn,Siegel:1984bm}. For a modern treatment from the superspace viewpoint on the constraints on $\cF_{AB}$, see \cite{Butter:2010jm}. To summarize, the function $\cF_{AB}(L)$ must satisfy the following four constraints in a superconformal setting
\bea
&\d_D \cF_{AB}(L) = - 3 \L_D \cF_{AB}(L),& \qquad  \cF_{ABC}{}^{ij}(L) = \cF_{A(BC)}{}^{ij} (L)\,, \nn\\
& \cF_{ABC}{}^{ij}(L)  L_{jk}^C  = - \ft12 \d^i{}_k \cF_{AB}(L),& \qquad \ve_{jk} \cF_{ABCD}{}^{ijkl}(L) = 0 \,.
\label{FABConstraints}
\eea
Any function $\cF_{AB}(L)$ that satisfies these constraints can be used to describe the couplings of $N$-number of linear multiplets. With these constraints and the composite expression for the $\l_{iA}$ we can now proceed to find how $Y_{ij A}$ and $F_{\m\n A}$ are expressed in terms of the components of the linear multiplet. Using the transformation rules (\ref{RigidConformalVector}), we find the following expressions for the composite $Y_{ij A}$ and $F_{\m\n A}$
\bea
Y_{ij A} &=& \cF_{AB} \Box L_{ij}^B + \cF_{ABC ij} N^B N^C + \cF_{ABC ij} E_a^B E^{a C} + \cF_{ABC}{}^{km} \partial_a L_{k(i}^B \partial^a L_{j)m}^{C} \nn\\
&& + 2 \cF_{ABC (i}{}^k E_a^B \partial^a L_{j)k}^C - \rmi \cF_{ABCD ij}{}^{kl} N^D \bar{\vf}_k^C \vf_l^B + \cF_{ABCD ij}{}^{kl} \bar{\vf}_k^B \slashed{E}^D \vf^C_l \nn\\
&& - 2 \cF_{ABC k (i} \bar{\vf}^{kB} \slashed\partial \vf_{j)}^C - \cF_{ABCD k (i}{}^{lm} \bar{\vf}^{k B} \slashed\partial L_{j)l}^L \vf_m^C  - \ft12 \cF_{ABCDEijmn}{}^{kl}\bar\vf_k^B \vf_l^C \bar{\vf}^{m D} \vf^{nE} \,,\nn\\
F_{\m\n A} &=& 4 \partial_{[\m} (\cF_{AB} E_{\n]}^B) + 2 \cF_{ABCD k}{}^l  \partial_{[\m} L^{kp B} \partial_{\n]} L_{lp}^C + 2 \partial_{[\m} (\cF_{ABCD ij} \bar{\vf}^{iB} \g_{\n]} \vf^{jC}) \,.
\label{CompositeYF}
\eea
We can now use the these expressions and the rigid density formulae (\ref{RigidConformalVector}) to provide an action for the rigid conformal linear multiplets. The bosonic part of the action reads
\bea
\cL &=& \cF_{AB} L_{ij}^A \Box L^{ij B} + \cF_{ABC}{}^{km} L^{ijA} \partial_a L_{ki}^B \partial^a L_{jm}^C + 2 \cF_{AB} N^A N^B  \nn\\
&& + \cF_{ABCij} L^{ij A} N^B N^C + 2 \cF_{AB} E_a^A E^{a B} + \cF_{ABC ij}   L^{ijA} E_a^B E^{a C}   \nn\\
&& + 2 \cF_{ABCi}{}^k  L^{ij A} E_a^B \partial^a L_{jk}^C + \cF_{ABC k}{}^l E^{ab A} \partial_a L^{kp B} \partial_b L_{lp}^C \,,
\label{ConformalLinearMultipletAction1}
\eea
and the fermionic part can simply be read from the composite expressions. As mentioned, there is no particular symmetry in the indices of $\cF_{AB}$, and indeed one can chose $\cF_{AB}$ to satisfy the constraints (\ref{FABConstraints}) that has no symmetry in $AB$ indices. One particular choice is to consider two linear multiplets, $(L_1^{ij}, \vf_1^i, E_1^a, N_1)$ and $(L_2^{ij}, \vf_2^i, E_2^a, N_2)$, and choose
\bea
\cF_{11} = L_2^{-1} \,, \qquad \cF_{12} = - L_2^{-3} L_{2pq} L_1^{pq} \,, \qquad \cF_{21} = \cF_{22} = 0 \,. 
\eea
This choice, which we will discuss in the next sections, is not symmetric in $\it (1,2)$ indices, and yet satisfies the constraints (\ref{FABConstraints}). Furthermore, in our construction above, the $A$ index is fixed thus it differs from the $B$ index that is being summed over.  However, if we sum over the $A$ index as well, then the action (\ref{ConformalLinearMultipletAction1}) can be simplified considerably. An example of such a choice is the non-interacting $N$ number of linear multiplets
\bea
\cF_{AB} = L^{-1} \d_{AB} \,.
\eea
When there is a sum over $A$ index, the action, including the fermionic terms, is given by
\bea
\cL &=& -\ft12 F_{AB} \partial_a L_{ij}^A \partial^a L^{ij B} + F_{AB} (N^A N^B + E_a^A E^{a B})  + F_{ABCk}{}^l E^{ab A} \partial_a L^{kp B} \partial_b L_{lp}^C \nn\\
&& - F_{AB} \bar{\vf}^A \slashed\partial \vf^B- F_{ABC}{}^{ij} N^A \bar{\vf}_i^B \vf_j^C + F_{ABC}{}^{ij} \bar{\vf}_i^B \slashed{E}^A \vf^C_j \nn\\
&& - \ft12 F_{ABCD}{}^{ijkl} \bar{\vf}_i^A \vf_j^B \bar{\vf}_k^C \vf_l^D  \,,
\label{SimplifiedRigid}
\eea
up to partial integrations. Here $F_{AB}$ and its descendants defined as \cite{deWit:2006gn}
\bea
F_{AB} &=& 2 \cF_{(AB)} + \cF_{ABC}{}^{ij} L_{ij}^A \,,\nn\\
F_{ABC}{}^{ij} &=& 3 \cF_{(ABC)}{}^{ij}  + \cF_{DABC}{}^{ijkl} L^D_{ij} \,,\nn\\
F_{ABCD}{}^{ijkl} &=& 4 \cF_{(ABCD)}{}^{ijkl} + \cF_{EABCD}{}^{ijklmn} L_{mn}^E \,.
\label{cyclicF}
\eea
This simplification also requires the use of following identities \cite{deWit:2006gn}
\bea
\cF_{ABCij} L_{kl}^B L^{kl C} &=& - \cF_{AB} L_{ij}^B \,,\nn\\
K_{ik} L^{jk} + K^{jk} L_{ik} &=& \d_i{}^k K^{kl} L_{kl} \,,\nn\\
K_{ij} L_{kl} - K_{kl} L_{ij} &=& \ve_{ik} \ve^{mn} (K_{lm} L_{nj} + K_{jm} L_{nl})|_{(i,j)(k,l)} \,, 
\eea
where $(i,j)$ and $(k,l)$ indicates symmetrization in the respected indices.

The same procedure can be repeated to construct an action for the vector multiplets, but this time we need to assume that the components of the linear multiplet can be written in terms of the elements of the vector multiplet
\bea
L_{ij A} &=& \cC_{ABC} \Big( 2 \r^B Y_{ij}^C - \ft12 \rmi\bar{\l}_{(i}^B \l_{j)}^C \Big) \,,\nn\\
\vf_{i\, A} &=& \cC_{ABC} \Big( \rmi \r^B \slashed{\partial}\l_i^C - \ft14 \g \cdot F^B \l_i^C + \ft12 \slashed\partial \r^B \l_i^C - Y_{ij}^B \l^{j C} \Big) \,,\nn\\
E^a_A &=& \cC_{ABC} \Big[ \partial_b \Big( - \r^B F^{ab\, C}  -	 \ft14 \rmi \bar{\l}^B \g^{ab} \l^C  \Big) - \ft18 \e^{abcde} F^B_{bc} F^C_{de} \Big] \,,\nn\\
N_A &=& \cC_{ABC} \Big( \r^B \Box \r^C + \ft12 \partial_\m \r^B \partial^\m \r^C - \ft14 F_{ab}^B F^{ab\, C} + Y_{ij}^B Y^{ij\, C} - \ft12 \bar{\l}^B \slashed\partial \l^C \Big) \,,
\label{VectorConformalEmbedding}
\eea
where $\cC_{ABC}$ is a constant symmetric tensor that determines the couplings of vector multiplets with each other. Using this composite formulae in the density formulae (\ref{RigidDensityFormulae}), an action for the conformal vector multiplets is given by \cite{Kugo:2000hn,Bergshoeff:2002qk,Kugo:2000af}
\bea
\cL &=& \cC_{ABC} \Bigg[ \Big( -\ft14 F_{\m\n}^A F^{\m\n\, B}  - \ft12 \bar{\l}^A \slashed{\partial} \l^B - \ft12 \partial_\m \r^A \partial^\m \r^B + Y_{ij}^A Y^{ij\,B}  \Big) \r^C   \nn\\
&& \qquad \qquad -\ft18 \rmi \bar{\l}^A \g \cdot F^B \l^C - \ft12 \rmi \bar{\l}^{i A} \l^{jB} Y_{ij}^C - \ft1{24}  \e^{\m\n\r\s\l} A_\m^A F_{\n\r}^B F_{\s\l}^C \Bigg] \,,
\label{ConformalRigidVectorAction}
\eea
up to total derivatives. 

%%%%%%%%%%%%%%%%%%%%%%%%%%%%%%%%%%%%
\section{Superconformal Linear Multiplets and Conformal Supergravity}{\label{Section2}}
%%%%%%%%%%%%%%%%%%%%%%%%%%%%%%%%%%%%

The superconformal realization of the vector and the linear multiplets in the rigid Minkowski background of the previous section can be generalized to a superconformal background. In this case, the superconformal transformations are gauged, and we need to introduce spacetime dependent transformation parameters along with the corresponding gauge fields. These gauge fields would form the so-called standard Weyl multiplet, which we give a detailed description in Appendix \ref{Appendix1}. In this case, the transformation rules for the linear multiplet reads \cite{Bergshoeff:2002qk}
\bea
\d L^{ij} &=& \rmi \bar{\e}^{(i} \vf^{j)} + 3 \L_D L^{ij} - 2 \L^{(i}{}_k L^{j)k} \,,\nn\\
\d \vf^{i} &=&  -\ft12 \rmi \slashed{\cD} L^{ij} \e_j - \ft12 \rmi \g^a E_a  \e^i + \ft12 N \e^i - \g \cdot T L^{ij} \e_j + 3 L^{ij}\eta_j + \ft72 \L_D \vf^i - \L^i{}_j \vf^j  \,,\nn\\
\d E_a &=&  -\ft12 \rmi \bar{\e} \g_{ab} \cD^b \vf + 2 \bar{\e} \g^b \vf T_{ab} - 2 \bar{\eta} \g_a \vf  + 4 \L_D E_a   \,,\nn\\
\d N &=& \ft12 \bar{\e} \slashed{\cD} \vf + \ft32 \rmi \bar{\e} \g \cdot T \vf   + \ft32 \rmi \bar{\eta} \vf  + 4 \L_D N     \,,
\label{SuperconformalLinear}
\eea
where the closure of the superconformal algebra on the components of the linear multiplet now requires
\bea
\cD^a E_a = 0 \,,
\eea 
which implies that $E_a$ can be solved as
\bea
E^a &=& - \ft1{12} e_\m{}^a \e^{\m\n\r\s\l} \cD_\n E_{\r\s\l} \,.
\label{EintermsofEmnr}
\eea
We can also define a 2-form potential $E_{\m\n}$ such as
\bea
E^a = e_\m{}^a \cD_\n E^{\m\n}  \quad \text{and} \quad  E_{\m\n\r} = \e_{\m\n\r\s\l} E^{\s\l}\,.
\label{defEmn}
\eea
Here, $E_{\m\n}$ transforms as
\bea
\d E^{\m\n} &=& - \ft12 \rmi \bar{\e} \g^{\m\n} \vf - \ft12 \bar{\p}_\r^i \g^{\m\n\r}\e^j L_{ij} \,,
\eea
and the covariant derivative of $E_{\m\n}$ in (\ref{defEmn}) is given by
\bea
\cD_\n E^{\m\n} &=& \partial_\n E^{\m\n} + \ft12 \rmi \bar{\p}_\n \g^{\m\n} \vf + \ft14 \bar{\p}_\r^i \g^{\m\n\r} \p_\n^j L_{ij} \,.
\eea
The supercovariant curvatures that we have used are defined as
\bea
\cD_\m L^{ij} &=& \partial_\m L^{ij} - 3 b_\m L^{ij} + 2 V_\m{}^{(i}{}_k L^{j)k} - \rmi \bar{\p}_\m^{(i} \vf^{j)} \,,\nn\\
\cD_\m \vf^i &=& \partial_\m \vf^i - \ft72 b_\m \vf^i + \ft14 \o_\m{}^{ab} \g_{ab} \vf^i - V_\m{}^{ij} \vf_j + \ft12 \slashed\cD L^{ij} \p_{\m j} + \ft12 \rmi \slashed{E} \p_\m^i \nn\\
&& -\ft12 N \p_\m^i + \g \cdot T L^{ij} \p_{\m j} - 3 L^{ij} \f_{\m j} \,,\nn\\
\cD_\m E_a &=& \partial_\m E_a - 4 b_\m E_a + \o_{\m ab} E^b + \ft12 \rmi \bar{\p}_\m \g_{ab} \cD^b \vf -2 \bar{\p}_\m \g^b \vf T_{ab} + 2 \bar{\phi}_\m \g_a \vf\,.
\label{LinearCovariantDerivatives}
\eea

The quantities that do not belong to the linear multiplet in the transformation rules and the covariant quantities, which we describe in Appendix \ref{Appendix1}, are the elements of the standard Weyl multiplet: $e_\m{}^a$ is the f\"unfbein, $b_\m$ is the gauge field that gauges dilatations, $V_\m{}^{ij}$ is the $\SU(2)$ R-symmetry gauge field, $\o_\m{}^{ab}$ is the spin connection, $f_\m{}^a$ is the gauge field for the special conformal transformations, $\p_\m^i$ is the $Q-$supersymmetry gauge field, and $\f_\m^i$ is the gauge field for $S-$supersymmetry. In addition to the gauge fields the standard Weyl multiplet consists of matter field: a real scalar $D$, an anti-symmetric tensor $T_{ab}$ and a symplectic Marojana spinor $\chi^i$.

For the vector multiplet, the transformation rules read \cite{Bergshoeff:2002qk}
\bea
\d \r &=& \ft12 \rmi \bar{\e} \l  + \L_D \r \,,\nn\\
\d A_\m &=&  -\ft12 \rmi \r \bar{\e} \p_\m + \ft12 \bar{\e} \g_\m \l  \,,\nn\\
\d \l^{i} &=& -\ft14 \g \cdot \widehat{F} \e^i - \ft12 \slashed{\cD} \r \, \e^i - Y^{ij} \e_j +\r \g \cdot T \e^i  + \r \eta^i  + \ft32 \L_D \l^i - \L^i{}_j \l^j    \,,\nn\\
\d Y^{ij} &=& \-\ft12 \bar{\e}^{(i} \slashed{\cD} \l^{j)} +\ft12 \bar{\e}^{(i} \g \cdot T \l^{j)} - 4 \rmi \r \bar{\e}^{(i}  \chi^{j)} + \ft12 \bar{\eta}^{(i} \l^{j)} + 2 \L_D Y^{ij} - 2 \L^{(i}{}_k Y^{j)k}  \,,
\label{SuperconformalVector}
\eea
where the superconformal field strength $\widehat{F}_{\m\n}^A$ is defined as
\bea
\widehat{F}_{\m\n} &=& 2 \partial_{[\m} A_{\n]} - \bar{\p}_{[\m} \g_{\n]} \l + \ft12 \rmi \r \bar{\p}_{[\m} \p_{\n]} \,,
\eea
and the supercovariant derivatives are given by
\bea
\cD_\m \r &=& \partial_\m \r - b_\m \r - \ft12 \rmi \bar{\p}_\m \l \,,\nn\\
\cD_\m \l^i &=& \partial_\m \l^i - \ft32 b_\m \l^i + \ft14 \o_\m{}^{ab} \g_{ab} \l^i - V_\m{}^{ij} \l_j + \ft14 \g \cdot \widehat{F} \p_\m^i \nn\\
&& + \ft12 \rmi \slashed\cD \r \, \p_\m^i + Y^{ij} \p_{\m j} - \r \g \cdot T \p_\m^i - \r \f_\m^i \,.
\eea

As in the rigid case, the vector and the linear multiplet actions can be constructed based on a density formulae, which is now given as \cite{Kugo:2000hn}
\bea
e^{-1} \cL &=& Y^{ij} L_{ij} + \rmi \bar{\l} \vf + \r N + A_a P^a   - \ft12 \bar{\p}_\m^i \g^\m \l^i L_{ij} + \ft12 \r \p_\m \g^\m \vf + \ft14 \rmi \r \bar{\p}_\m^i \g^{\m\n} \p_\n^j L_{ij}     \,,
\label{SuperconformalDensity}
\eea
where $P_a$ is the bosonic part of the superconformal curvature $E_a$
\bea
P^a &=&E^a + \ft12 \rmi \bar{\p}_b \g^{ba} \vf  + \ft14 \bar{\p}_b^i \g^{abc} \p_c^j L_{ij} \,.
\label{EintermsofP}
\eea
Using (\ref{EintermsofEmnr}), we can also express $P_a$ in terms of $E_{\m\n\r}$ as
\bea
P^a &=& - \ft{1}{12} e_\m{}^a \e^{\m\n\r\s\l} \partial_\n E_{\r\s\l} \,.
\eea
Just as in the rigid superconformal case, we can construct the vector multiplets out of linear multiplets, however this time they should be consistent with the transformation rules in which the transformation parameters are local. The ansatz for the $\r_A$ in the rigid case, (\ref{LowestLinear}) does not need a modification, and can be used as the starting point. Then, the composite vector multiplet, which can be found as a sequence of $Q-$supersymmetry transformations, is given by
\bea
\r_A &=& 2 \cF_{AB} N^B - \rmi \cF_{ABC}{}^{ij} \bar{\vf}_i^B \vf_j^C \,,\nn\\
\l_{i A} &=&  - 2\rmi \cF_{AB} \slashed\cD \vf_i^B + 2 \cF_{ABC ij} \vf^{jB} N^C + 2 \rmi \cF_{ABC ij} \slashed{E}^B \vf^{j C}  - 2 \rmi \cF_{ABC}{}^{jk} \slashed\cD  L_{ij}^B \vf_k^C \nn\\
&& + 16 \cF_{AB} L_{ij}^B \chi^j + 4 \cF_{AB} \g \cdot T \vf^B_i - 2 \rmi \cF_{ABCDij}{}^{kl} \vf^{j D} \bar{\vf}_k^B \vf_l^C \,,\nn\\
Y_{ij A} &=& \cF_{AB} \Box^c L_{ij}^B + \cF_{ABC ij} N^B N^C + \cF_{ABC ij} E_a^B E^{a C} + \cF_{ABC}{}^{km} \cD_a L_{k(i}^B \cD^a L_{j)m}^{C} \nn\\
&& + 2 \cF_{ABC (i}{}^k E_a^B \cD^a L_{j)k}^C + \ft83 \cF_{AB} L_{ij}^A T^2 + 4 \cF_{AB} L_{ij}^B D- \rmi \cF_{ABCD ij}{}^{kl} N^D \bar{\vf}_k^C \vf_l^B   \nn\\
&& + \cF_{ABCD ij}{}^{kl} \bar{\vf}_k^B \slashed{E}^D \vf^C_l - 16 \rmi \cF_{AB} \bar{\chi}_{(i} \vf_{j)}^B - 16 \rmi \cF_{ABCij} L_{kl}^B \bar{\chi}^k \vf^{l C} \nn\\
&& - 2 \cF_{ABC k (i} \bar{\vf}^{kB} \slashed\cD \vf_{j)}^C - \cF_{ABCD k (i}{}^{lm} \bar{\vf}^{k B} \slashed\cD L_{j)l}^L \vf_m^C - 2 \rmi \cF_{ABCij} \bar{\vf}^B \g \cdot T \vf^C \nn\\
&& - \ft12 \cF_{ABCDEijmn}{}^{kl}\bar\vf_k^B \vf_l^C \bar{\vf}^{m D} \vf^{nE} \,,\nn\\
\widehat{F}_{\m\n A} &=& 4 \cD_{[\m} (\cF_{AB} E	_{\n]}^B) + 2 \cF_{AB} \widehat{R}_{\m\n}{}^{ij}(V) L_{ij}^B + 2 \cF_{ABCD k}{}^l  \cD_{[\m} L^{kp B} \cD_{\n]} L_{lp}^C  \nn\\
&& + 2 \cD_{[\m} (\cF_{ABCD ij} \bar{\vf}^{iB} \g_{\n]} \vf^{jC}) - \rmi \cF_{AB} \bar{\vf}^B \widehat{R}_{\m\n}(Q) \,,
\label{VectorCompositeSUGRA}
\eea
where $\widehat{R}_{\m\n}{}^{ij}(V)$ and $\widehat{R}_{\m\n}(Q)$ are the supercovariant curvatures of the standard Weyl multiplet fields $V_\m{}^{ij}$ and $\p_\m^i$ respectively. For a single linear multiplet, this result matches with \cite{Coomans:2012cf}. The superconformal d'Alembertian of $L_{ij}$ is defined as
\bea
\Box^c L^{ij}_A &=& (\partial^a - 4 b^a + \o_b{}^{ba}) \cD_a L^{ij}_A + 2 V_a^{(i}{}_k \cD^a L^{j)k}_A + 6 f_a^a L^{ij}_A - \rmi \bar{\p}^{a(i} \cD_a \vf^{j)}_A \nn\\
&&  + 4 \bar{\p}^{a{(i}}\g_a \chi^{k)} L^j{}_{kA} + 4 \bar{\p}^{a{(j}}\g_a \chi^{k)} L^i{}_{kA}- 6 L^{ij}_A \bar{\p}^a \g_a \chi \nn\\
&& - \bar{\vf}_A^{(i} \g \cdot T \g^a \p_a^{j)} + \bar{\vf}^{(i}_A \g^a \phi^{j)}_a \,.
\label{BoxL}
\eea
Although the form of composite $\widehat{F}_{ab A}$ is manifestly covariant, it is useful to write down the bosonic part of $\widehat{F}_{ab A}$ in a different way, which is useful when we use the composite expressions to form an action
\bea
F_{\m\n A} &=&  4 \partial_{[\m} (\cF_{AB} E_{\n]}^B +  \cF_{AB} V_{\n]}^{ij} L^B_{ij})  + 2 \cF_{ABC k}{}^l  \partial_{[\m} L^{kp B} \partial_{\n]} L_{lp}^C \,.
\label{FAB}
\eea
Consequently, the action that describes the supergravity coupling of $N$-number of linear multiplets is given by
\bea
e^{-1} \cL &=& \cF_{AB} L_{ij}^A \Box^c L^{ij B} + \cF_{ABC}{}^{km} L^{ijA} \cD_a L_{ki}^B \cD^a L_{jm}^C + 2 \cF_{AB} N^A N^B  \nn\\
&& + \cF_{ABCij} L^{ij A} N^B N^C + 2 \cF_{AB} E_a^A E^{a B} + \cF_{ABC ij}   L^{ijA} E_a^B E^{a C}   \nn\\
&& + 2 \cF_{ABCi}{}^k  L^{ij A} E_a^B \cD^a L_{jk}^C + \cF_{ABC k}{}^l E^{ab A} \partial_a L^{kp B} \partial_b L_{lp}^C \nn\\
&& + \ft83 \cF_{AB} L^{ij A} L_{ij}^B T^2 + 4 \cF_{AB} L^{ij A} L_{ij}^B D + 2 \cF_{AB} E_a^A L_{ij}^B V^{a ij} \,.
\label{LinearConformalAction}
\eea
Here, we only provide the bosonic part of the action, and the fermionic part can be read from the composite formulae. Assuming that there is a summation over the $A$ index, and using $F_{AB}$ and its descendants (\ref{cyclicF}), the bosonic part of the linear multiplet action can be simplified to
\bea
e^{-1} \cL &=& - \ft38 F_{AB} L_{ij}^A L^{ij B} R + 4  F_{AB} L_{ij}^A L^{ij B} D + \ft83 F_{AB} L_{ij}^A L^{ij B} T^2  - \ft12 F_{AB} D_\m L_{ij}^A D^\m L^{ij B} \nn\\
&&  + F_{AB}(N^A N^B + E_\m^A E^{\m B}) + F_{ABCk}{}^l E^{\m\n} \partial_\m L^{kpB} \partial_\n L_{lp}^C + 2 F_{AB} E_\m^A L_{ij}^B V^{\m ij} \,,
\eea
where we have defined the $\SU(2)$ covariant derivative
\bea
D_\m L^{ij A} &=& \partial_\m L^{ij A}  + 2 V_\m{}^{(i}{}_k L^{j)k A} \,,
\eea
and used $E^a = \cD^b E_{ab}$ and $\cF_{ABCDi[jk]l} = 0$.  We have also used the definitions of the superconformal covariant quantities given in (\ref{LinearCovariantDerivatives}), (\ref{EintermsofP}) and (\ref{BoxL}). Note that we did not fix the bosonic field $b_\m = 0$, however it does not show up in the Lagrangian. This is because the $b_\m$ terms in the superconformal d'Alembertian $\Box^c L_{ij}$ precisely cancels the $b_\m$ terms in the supercovariant curvature of $L_{ij}$ (\ref{LinearCovariantDerivatives}), and the remaining terms are the Ricci scalar, $R$, that is associated with the gravitational field, as well as the $\SU(2)$ covariant derivative of $L_{ij}$. This is a typical property of conformal gravity dictated by the special conformal invariance.

For the construction of a two-derivative vector multiplet action, we construct the elements of the linear multiplet in terms of the vector multiplet fields  \cite{Fujita:2001kv,Bergshoeff:2002qk}
\bea
L_{ij A} &=& \cC_{ABC} \Big( 2 \r^B Y_{ij}^C - \ft12 \rmi \bar{\l}_{(i}^B \l_{j)}^C \Big) \,,\nn\\
\vf_{i\, A} &=& C_{ABC} \Big( \rmi \r^B \slashed{\cD}\l_i^C + 2 \r^B \g \cdot T \l_i^C - 8 \r^B \r^C \chi_i - \ft14 \g \cdot \widehat{F}^B \l_i^C + \ft12 \slashed\cD \r^B \l_i^C - Y_{ij}^B \l^{j C} \Big) \,,\nn\\
E^a_A &=& \cC_{ABC} \Big[ \cD_b \Big( - \r^B \widehat{F}^{ab\, C} + 8 \r^B \r^C T^{ab} - \ft14 \rmi \bar{\l}^B \g^{ab} \l^C  \Big) - \ft18 \e^{abcde} \widehat{F}_{bc}^B \widehat{F}_{de}^C \Big] \,,\nn\\
N_A &=& \cC_{ABC} \Big[ \r^B \Box^c \r^C + \ft12 \cD_\m \r^B \cD^\m \r^C - \ft14 \widehat{F}_{ab}^B \widehat{F}^{ab\, C} + Y_{ij}^B Y^{ij\, C} + 8 \r^B \widehat{F}_{ab}^C T^{ab} \nn\\
&& \qquad - 4 \r^B \r^C \Big(D + \ft{26}{3} T^2\Big) - \ft12 \bar{\l}^B \slashed\cD \l^C + \rmi \bar{\l}^B \g \cdot T \l^C + 16 \rmi \r^B \bar\chi \l^C \Big] \,,
\label{VectorSuperconformalEmbedding}
\eea
where the superconformal d'Alembertian is defined as
\bea
\Box^c \r^A &=& (\partial^a - 2 b^a + \o_b{}^{ba}) \cD_a \r^A - \ft12 \rmi \bar{\p}_a \cD^a \l^A - 2 \r^A \bar{\p}_a \g^a \chi \nn\\
&& + \ft12 \bar{\p}_a \g^a \g \cdot T \l^A + \ft12 \bar{\f}_a \g^a \l^A + 2 f_a^a \r^A \,.
\eea
Using the composite expressions and the density formulae (\ref{SuperconformalDensity}), we can describe the supergravity coupling of $N$-number of vector multiplets as
\bea
e^{-1} \cL &=& \cC_{ABC} \Big[ -\ft1{24} \r^A \r^B \r^C R - \ft14 \r^A F_{\m\n}^B F^{\m\n C} -\ft12 \r^A \partial_\m \r^B \partial^\m \r^C + \r^B Y_{ij}^A Y^{ij C} \nn\\
&& \qquad - \ft43 \r^A \r^B \r^C \Big( D + \ft{26}{3} T^2 \Big) + 4 \r^A \r^B F_{\m\n}^C T^{\m\n} - \ft1{24} \e^{\m\n\r\s\l} A_\m^A F_{\n\r}^B F_{\s\l}^C \Big] \,.
\label{VectorConformalAction}
\eea
Here, we again provide the bosonic part of the action, and the fermionic part can be read from the composite formulae.

%%%%%%%%%%%%%%%%%%%%%%%%%%%%%%%%%%%%
\section{Poincar\'e Supergravity and Off-Diagonal Invariants}{\label{Section3}}
%%%%%%%%%%%%%%%%%%%%%%%%%%%%%%%%%%%%

In the previous sections, we review the construction of the supergravity coupling of vector multiplets, and constructed an action for $N$-number of local superconformal linear multiplets. In the following subsections, we will consider the byproducts of our constructions, i.e. the off-shell Poincar\'e supergravity and higher derivative invariants.

%%%%%%%%%%%%%%%%%%%%%%%%%%%%%%%%%%%%
\subsection{Poincar\'e Supergravity}
%%%%%%%%%%%%%%%%%%%%%%%%%%%%%%%%%%%%

There are more than one way to obtain an off-shell Poincar\'e supergravity using linear, vector and Weyl multiplets. In Table \ref{Table1}, we provide a list of possible constructions of an off-shell supergravity using these multiplets. In this paper, we will not use the dilaton Weyl multiplet, therefore, will discuss the details here. The constructions based on the dilaton Weyl multiplet can be found in \cite{Ozkan:2013nwa,Coomans:2012cf,Ozkan:2013uk}, and the dilaton Weyl multiplet itself is briefly discussed in Appendix \ref{Appendix1}.

\begin{table}[h!]
\centering
\begin{tabular}{cccc}
\hline
\multicolumn{1}{|c|}{\textbf{Weyl Multiplet}}                                                                                                                  & \multicolumn{1}{c|}{\textbf{Compensator(s)}}                                                                                                                & \multicolumn{1}{c|}{\textbf{Gauge Fixing}}                                                                                  & \multicolumn{1}{c|}{\textbf{\begin{tabular}[c]{@{}c@{}}Supergravity \\ Multiplet\end{tabular}}}                                                                          \\ \hline
\multicolumn{1}{|c|}{\begin{tabular}[c]{@{}c@{}}Standard Weyl Multiplet\\ $(e_\m{}^a, \p_\m^i, V_\m^{ij}, b_\m, D, T_{ab}, \chi^i)$\end{tabular}}              & \multicolumn{1}{c|}{\begin{tabular}[c]{@{}c@{}}Linear Multiplet\\ $(L^{ij}, \vf^i, E_a, N)$\\ Vector Multiplet\\ $(\rho, A_\m, \l_i, Y_{ij})$\end{tabular}} & \multicolumn{1}{c|}{\begin{tabular}[c]{@{}c@{}}$b_\m = 0$\\ $L_{ij} = \ft1{\sqrt2} \d_{ij} L$\\ $L=1$\\ $\vf_i = 0$\end{tabular}} & \multicolumn{1}{c|}{\begin{tabular}[c]{@{}c@{}}$(e_\m{}^a, V_\m, V^\prime_\m{}^{ij}, D,$ \\ $T_{ab}, 
	\r, A_\m, Y_{ij}, E_a, N, $\\ $\p_\m^i, \l^i, \chi^i)$,\end{tabular}} \\ \hline
\multicolumn{1}{|c|}{\begin{tabular}[c]{@{}c@{}}Standard Weyl Multiplet\\ $(e_\m{}^a, \p_\m^i, V_\m^{ij}, b_\m, D, T_{ab}, \chi^i)$\end{tabular}}              & \multicolumn{1}{c|}{\begin{tabular}[c]{@{}c@{}}Linear Multiplet\\ $(L^{ij}, \vf^i, E_a, N)$\\ Vector Multiplet\\ $(\rho, A_\m, \l_i, Y_{ij})$\end{tabular}} & \multicolumn{1}{c|}{\begin{tabular}[c]{@{}c@{}}$b_\m = 0$\\ $\r = 1$\\ $\l_i= 0$\end{tabular}}                              & \multicolumn{1}{c|}{\begin{tabular}[c]{@{}c@{}}$(e_\m{}^a, V_\m{}^{ij}, D, T_{ab}, $\\ $L_{ij}, A_\m,Y_{ij}, E_a, N,$\\ $\p_\m^i, \vf^i, \chi^i)$,\end{tabular}}         \\ \hline
\multicolumn{1}{|c|}{\begin{tabular}[c]{@{}c@{}}Dilaton Weyl Multiplet\\ $(e_\m{}^a, \p_\m^i, V_\m^{ij}, b_\m, \sigma, C_{\m}, B_{\m\n} \psi^i)$\end{tabular}} & \multicolumn{1}{c|}{\begin{tabular}[c]{@{}c@{}}Linear Multiplet\\ $(L^{ij}, \vf^i, E_a, N)$\end{tabular}}                                                   & \multicolumn{1}{c|}{\begin{tabular}[c]{@{}c@{}}$b_\m = 0$\\ $L_{ij} = \ft1{\sqrt2} \d_{ij} L$\\ $L=1$\\ $\vf_i = 0$\end{tabular}} & \multicolumn{1}{c|}{\begin{tabular}[c]{@{}c@{}}$(e_\m{}^a, V_\m, V^\prime_\m{}^{ij}, \s, $\\ $B_{\m\n}, C_\m, E_a, N,$\\ $ \p_\m^i, \p^i)$,\end{tabular}}                  \\ \hline
\multicolumn{1}{|c|}{\begin{tabular}[c]{@{}c@{}}Dilaton Weyl Multiplet\\ $(e_\m{}^a, \p_\m^i, V_\m^{ij}, b_\m, \sigma, C_{\m}, B_{\m\n} \psi^i)$\end{tabular}} & \multicolumn{1}{c|}{\begin{tabular}[c]{@{}c@{}}Linear Multiplet\\ $(L^{ij}, \vf^i, E_a, N)$\end{tabular}}                                                   & \multicolumn{1}{c|}{\begin{tabular}[c]{@{}c@{}}$b_\m = 0$\\ $\s = 1$\\ $\p_i= 0$\end{tabular}}                              & \multicolumn{1}{c|}{\begin{tabular}[c]{@{}c@{}}$(e_\m{}^a, V_\m{}^{ij}, B_{\m\n}, C_\m,$\\ $L_{ij},E_a, N, $\\ $\p_\m^i, \vf^i)$,\end{tabular}}                          \\ \hline
\\
\end{tabular}
\caption{A list of possible constructions of off-shell supergravity models using standard/dilaton Weyl multiplet, a single vector multiplet and a single linear multiplet. Gauge fixing by use of multiple multiplets correspond to an appropriate combination of these gauge fixing choices. The gauge choice $L_{ij} = 1/\sqrt{2} \d_{ij} L$ is not a must, and does not fix a conformal symmetry but breaks the R-symmetry group $\SU(2)_R$ to $\U(1)_R$, which simplifies calculations tremendously. When the standard Weyl multiplet is used, both the linear and the vector multiplets must be utilized for the construction of a supergravity theory, and the difference between the off-shell supergravity multiplets lie in the different choices for the gauge fixing. When the dilaton Weyl multiplet is used, the linear multiplet itself is sufficient for the construction of an off-shell Poincar\'e supergravity.}
\label{Table1}
\end{table}

As mentioned, the construction of a supergravity based on a standard Weyl multiplet requires the use of both vector and the linear multiplets. Thus, our starting point for the construction of a supergravity is a combination of the vector multiplet action (\ref{VectorConformalAction}) and the linear multiplet action (\ref{LinearConformalAction}). We let this action be $\cL = - \cL_L - 3 \cL_V$, which reads
\bea
e^{-1} \cL &=& \ft18 ( \cC + 3 F_{AB} L_{ij}^A L^{ij B} ) R + 4 ( \cC - F_{AB} L_{ij}^A L^{ij B} ) D + \ft13 (104 \cC - 8 F_{AB} L_{ij}^A L^{ij B} ) T^2 \nn\\
&& + \ft12 F_{AB} D_\m L_{ij}^A D^\m L^{ij B}  - F_{AB}N^A N^B - F_{AB} E_\m^A E^{\m B} - F_{ABCk}{}^l E^{\m\n} \partial_\m L^{kpB} \partial_\n L_{lp}^C \nn\\
&& - 2 F_{AB} E_\m^A L_{ij}^B V^{\m ij} + \ft34 C_{ABC} \r^A F_{\m\n}^B F^{\m\n C} + \ft32 C_{ABC} \r^A \partial_\m \r^B \partial^\m \r^C \nn\\
&&  - 3 C_{ABC} \r^A Y_{ij}^B Y^{ij C} - 12 C_{ABC} \r^A \r^B F_{\m\n}^C T^{\m\n} + \ft18 C_{ABC} \e^{\m\n\r\s\l} A_\m^A F_{\m\n}^B F_{\r\s}^C \,,
\label{ConformalSugra}
\eea
where we have defined $\cC \equiv \cC_{ABC} \r^A \r^B \r^C$. 

The Lagrangian (\ref{ConformalSugra}) is invariant under the full superconformal group, and the conformal symmetries must be fixed to have an off-shell Poincar\'e supergravity. The canonical Einstein-Hilbert term can be recovered by the gauge choice
\bea
\cC + 3 F_{AB} L_{ij}^A L^{ij B} &=& 4 \,,
\eea
which would fix the dilatations. Consequently, the gauge choices
\bea
\ft12 \cC_{ABC} \r^A \r^B \l_i^C + 2 F_{AB} L_{ij}^A \vf^{j B} = 0 \,, \qquad b_\m = 0 \,,
\eea
would fix $S-$supersymmetry and special conformal transformations respectively. The resulting model would then describe the off-shell supergravity coupled to vector and linear multiplets. For simplicity, let us focus on a single linear multiplet choice which would correspond to $\cF_{11} = L^{-1}$. Noticing that $D$ equation of motion is given by
\bea
0 &=& \cC - F_{AB} L_{ij}^A L^{ij B} \,,
\eea
it is sufficient to gauge fix using a single linear multiplet to obtain the canonical Einstein-Hilbert term in the on-shell theory. This can be done by the following gauge choice
\bea
L = 1 \,, \quad L_{ij} = \ft1{\sqrt2} \d_{ij}  \,, \quad \vf_i = 0 \,, \quad b_\m = 0 \,,
\label{GaugeFixing}
\eea
where the first choice would fix dilatations, the second choice would break the $\SU(2)$ R-symmetry group to $\U(1)$, the third choice would fix the S-supersymmetry transformations and the last choice would fix the special conformal transformations. Adapting the notation
\bea
\cC_A = 3 \, \cC_{ABC} \r^B \r^C \,, \quad \cC_{AB} = 6 \, \cC_{ABC} \r^C \,,
\eea
the off-shell Poincar\'e theory is given by \cite{Ozkan:2013nwa}
\bea
e^{-1} \cL &=& \ft18 (\cC+3)R + \ft13 (104 \cC - 8) T^2 + 4 (\cC-1)D - N^2- P_\m P^\m + V_\m^{'ij} V^{'\m}_{ij} - \sqrt2 V_\m P^\m \nn\\
&&+ \ft34 \cC_{ABC} \r^A F_{\m\n}^B F^{\m\n\, C}+ \ft32 C_{ABC} \r^A \partial_\m \r^B \partial^\m \r^C - 3 \cC_{ABC} \r^A Y_{ij}^B Y^{ij\,C}\nn\\
&& - 12 \cC_{ABC} \r^A \r^B F_{\m\n}^C T^{\m\n} + \ft18 \e^{\m\n\r\s\l} \cC_{ABC} A^A_\m F_{\n\r}^B F_{\s\l}^C \,.
\label{SWSUGRA}
\eea
where we have decomposed the $\SU(2)_R$ gauge field $V_\m^{ij}$ into its trace and the traceless parts
\bea
V_\m^{ij} = V_\m^{\prime ij} + \ft12 \d_{ij} V_\m \,, \quad V_\m^{\prime ij} \d_{ij} = 0 \,.
\eea
We can now eliminate the auxiliary fields $D, T_{ab}, P_\m, V_\m V_\m^{\prime ij}, N$ and $Y_{ij}$ by their field equations. As the field equation for $V_\m$ implies $P_\m = 0$, we notice that the auxiliary fields $P_\m, V_\m, V_\m^{ij}, N$ and $Y_{ij}$ are set to zero by their field equations. Then the $D$ and $T_{ab}$ equations give rise to
\bea
0 &=& \cC - 1 \,,\nn\\
0 &=& \ft23 (104\cC - 8) T_{ab} - 4 \cC_A F_{\m\n}^A \,.
\eea
Consequently, the on-shell action is given by
\bea
e^{-1} \cL &=& \ft12 R + \ft18 (\cC_{AB} - \cC_A \cC_B) F_{\m\n}^A F^{\m\n B} + \ft14 \cC_{AB} \partial_\m \r^A \, \partial^\m \r^B \nn\\
&& + \ft18 \e^{\m\n\r\s\l} \cC_{ABC} A^A_\m F_{\n\r}^B F_{\s\l}^C \,.
\eea
Here, the $\U(1)$ R-symmetry is ungauged, hence the maximally symmetric solution of the theory is given by Minkowski$_5$. The gauged model can be obtained by adding the density formulate (\ref{SuperconformalDensity}) to the conformal supergravity model (\ref{ConformalSugra}), gauge fixing according to (\ref{GaugeFixing}), and using the field equations of the corresponding theory. This procedure is presented in detail in \cite{Ozkan:2013nwa}.

%%%%%%%%%%%%%%%%%%%%%%%%%%%%%%%%%%%%
\subsection{Off-Diagonal Invariants}
%%%%%%%%%%%%%%%%%%%%%%%%%%%%%%%%%%%%

When the linear and the vector multiplets are used as compensators along with the standard Weyl multiplet, the construction of the Weyl squared and the Ricci scalar squared invariants are discussed in \cite{Ozkan:2013nwa,Hanaki:2006pj}. However, there can also be supersymmetric higher derivative models that consists both the curvature and the auxiliary scalar, i.e., $RN$, but not a pure curvature term. These are called the off-diagonal invariants. Such models already known to exist in lower dimensional supergravity theories, i.e. see \cite{Bergshoeff:2010mf,Bergshoeff:2014ida,Alkac:2014hwa} for the three dimensional $\cN = 1,2$ examples.

For the construction of higher derivative off-diagonal supergravity models, let's first review the composite formulae (\ref{VectorCompositeSUGRA}). For a single linear multiplet, which corresponds to $\cF_{11} = L^{-1}$, and with the gauge fixing conditions (\ref{GaugeFixing}), the bosonic part of the map between vector and linear multiplet is given by 
\bea
\underline{\r} &=& 2N,  \nn\\
\underline{Y}^{ij} &=&  \ft1{\sqrt2} \d^{ij}\Big(- \ft3{8} R -  N^2 - P^2 + \ft83 T^2 + 4D - V_{a}^{'kl} V_{kl}^{'a} \Big)   \nn\\
&& + 2 P^a V'_{a}{}^{ij}  - \sqrt2 \nabla^a V'_{a}{}^{m(i} \d^{j)}{}_{m} , \nn\\
\underline{F}^{\m\n} &=& 2\sqrt{2} \partial^{[\m} \Big( V^{\n]} + \sqrt2 P^{\n]} \Big) \,.
\label{fixedmap}
\eea
An important fact about the gauge fixed map (\ref{fixedmap}) is that for a the single multiplet choice $\cF_{11} = L^{-1}$, the composite expression for the $F_{\m\n}$ becomes (\ref{FAB}) 
	\bea
	F_{\m\n} &=& 4 \partial_{[\m} \left(L^{-1} E_{\n]} + L^{-1} V_{\n]}^{ij} L_{ij} \right) - 2 L^{-3} L_k{}^l \partial_{[\m} L^{kp} \partial_{\n]} L_{lp} \,.
	\eea
Imposing the gauge fixing conditions (\ref{GaugeFixing}), this equation reduces to the composite expression in (\ref{fixedmap}), thus, $F_{\m\n}$ becomes exact in the sense that it can be written as a curl of a quantity
\bea
A_\m &=& \sqrt2 (V_\m + \sqrt2 P_\m) \,.
\eea
From the vector multiplet action (\ref{VectorConformalAction}), it is evident that the number of $\r$'s that multiply $R$ determines the type of the off-diagonal invariant. Therefore, the first invariant we present here is the $RN^3$ invariant given by
\bea
e^{-1} \cL_{RN^3} &=& -\ft1{24} \underline{\r}^3 R - \ft14 \underline{\r} \underline{F}^{\m\n} \underline{F}_{\m\n} - \ft12 \underline{\r} \partial_\m \underline{\r} \partial^\m \underline{\r} + \underline{\r} \underline{Y}^{ij} \underline{Y}_{ij} \nn\\
&& -\ft43 \underline{\r}^3 (D + \ft{26}3 T^2 ) + 4 \underline{\r}^2 \underline{F}^{\m\n} T_{\m\n} -\ft1{24}\e_{\m\n\r\s\l} \underline{A}^\m \underline{F}^{\n\r} \underline{F}^{\s\l} \,.
\eea
This action can simply be obtained by choosing a single vector multiplet via $C_{111} = 1$ and using the gauge-fixed composite formulae (\ref{fixedmap}). When the composite expressions are expanded, the supersymmetric completion of the $RN^3$ action is given by
\bea
e^{-1} \cL_{RN^3} &=& \ft13 R N^3 + N G_{\m\n} G^{\m\n} + \ft{32}{2} N^3 \Big(D + \ft{26}3 T^2 \Big) - 16\sqrt2 N^2 G_{\m\n} T^{\m\n} \nn\\
&& + 4 N \partial_\m N \partial^\m N - 2N \Big( \ft38 R + N^2 + P^2 + \ft83 T^2 + 4D - 2 Z^\m \bar{Z}_\m \Big)^2\nn\\
&&  + 8 N | \nabla^\m Z_\m + \rmi \sqrt2 P^\m Z_\m |^2  + \ft1{6\sqrt2} \e^{\m\n\r\s\l} C_\m G_{\n\r} G_{\s\l} \,,
\eea
up to an overall minus sign. Here, we have defined
\bea
Z_\m = V_\m^{\prime 12} + \rmi V_\m^{\prime 11}\,, \qquad C_\m = V_\m + \sqrt2 P_\m \,, \qquad G_{\m\n} = \partial_\m C_\n - \partial_\n C_\m \,.
\eea
Another off-diagonal supersymmetric invariant, the supersymmetric completion of the $RN$ invariant can be constructed by setting $C_{1AB} = a_{AB}$ and all other possibilities to zero. In this case, we obtain
\bea
e^{-1} \cL_{RN} &=& a_{AB} \Big( \underline{\r} Y_{ij}^A Y^{ij B} + 2 \r^A Y_{ij}^B \underline{Y}^{ij} - \ft18 \r^A \r^B \underline{\r} R - \ft14 \underline{\r} F_{\m\n}^A {F}^{\m\n B} - \ft12 \r^A F_{\m\n}^B \underline{F}^{\m\n} \nn\\
&& + \ft12 \underline{\r} \partial_\m \r^A \partial^\m \r^B +  \underline{\r} \r^A \Box \r^B - 4 \r^A \r^B \underline{\r} \Big(D + \ft{26}3 T^2 \Big) + 4 \r^A \r^B \underline{F}^{\m\n} T_{\m\n} \nn\\
&& + 8 \underline{\r} \r^A F_{\m\n}^B T^{\m\n} - \ft18 \e^{\m\n\r\s} \underline{A}^\m F^{\n\r A} F^{\s\l B} \Big) \,.
\eea
Finally, when the composite expressions (\ref{fixedmap}) are used, the supersymmetric completion of the $RN$ invariant is given by
\bea
e^{-1} \cL_{RN} &=& a_{AB} \Big[ \ft14 \r^A \r^B N R + \ft1{2} N F_{\m\n}^A F^{\m\n B} + \ft1{\sqrt2} \r^A F_{\m\n}^B G^{\m\n} -  N \partial_\m \r^A \partial^\m \r^B - 2 N \r^A \Box \r^B \nn\\
&& - 2 N Y_{ij}^A Y^{ij B} + \sqrt2 \r^A Y_{ij}^B \d^{ij} \Big(\ft3{8} R +  N^2 + P^2 - \ft83 T^2 - 4D + V_{\m}^{'kl} V_{kl}^{'\m} \Big) \nn\\
&& - 2 \r^A Y_{ij}^B (2 P^\m V'_{\m}{}^{ij}  - \sqrt2 \nabla^\m V'_{\m}{}^{mi} \d^{j}{}_{m}) + 8 N \r^A \r^B (D + \ft{26}3 T^2 ) \nn\\
&& - 4\sqrt2 \r^A \r^B G_{\m\n} T^{\m\n} - 16 N \r^A F_{\m\n}^B T^{\m\n} + \ft1{4\sqrt2}\e^{\m\n\r\s} C^\m F^{\n\r A} F^{\s\l B} \Big] \,,
\eea
up to an overall minus signature. Note that one can choose $\a_{IJ} = \a_{11} = 1$ to truncate the model to a single vector multiplet. 

In principle, one expects to have an $RN^2$ action along with the $RN$ and $RN^3$ invariants. Unfortunately, this is not the case since the supersymmetric completion of the $RN^2$ also includes the $R^2$ itself, hence not giving rise to an off-diagonal supersymmetric action. This situation is very similar to the three dimensional $\cN = 2$ case where one has $RD$ and $RD^3$ invariants, but not $RD^2$ \cite{Alkac:2014hwa,Kuzenko:2015jda}, where $D$ is the auxiliary component of a three dimensional $\cN = 2$ vector multiplet. Finally, we note that using the composite expression for the vector multiplet (\ref{fixedmap}), it is possible to construct other off-diagonal invariants such as $N C_{\m\n\r\s} C^{\m\n\r\s}$ invariant where $C_{\m\n\r\s}$ is the Weyl tensor. Such an invariant can be easily produced using the Weyl tensor squared invariant \cite{Hanaki:2006pj} and assuming that the vector multiplet is a composite multiplet defined by the expressions (\ref{fixedmap}). 

%%%%%%%%%%%%%%%%%%%%%%%%%%%%%%%%%%%%
\subsection{Comments on the Higher Derivative Yang-Mills Action}
%%%%%%%%%%%%%%%%%%%%%%%%%%%%%%%%%%%%

In the previous section, we discuss the construction of higher derivative linear multiplet actions, which led us to higher derivative supergravity invariants. We can use the same mechanism to produce higher order supersymmetric invariants for vector multiplets. As the first step, the bosonic part of the composite formulae (\ref{VectorSuperconformalEmbedding}) for a single vector multiplet is given by
\bea
\underline{L}^{ij} &=& 2 \r Y^{ij} \,,\nn\\
\underline{N} &=& - \ft{1}{16} \r^2 R+  \r \Box \r + \ft12 \partial_\m \r \partial^\m \r - \ft14 F_{\m\n} F^{\m\n} + Y^{ij} Y_{ij}  + 8 F_{\m\n} T^{\m\n} - 4 \r^2 \Big( D + \ft{26}3 T^2 \Big) \,,\nn\\
\underline{P}^a &=& \partial_b \Big( - \r F^{ab} + 8 \r^2 T^{ab} \Big) - \ft18 \e^{abcde} F_{bc} F_{de} \,,
\label{CompositeLinear}
\eea
which can be obtained by setting $\cC_{111} = 1$. With these composite expressions, it is obvious that a conformal linear multiplet action with an $N^2$ term would produce a higher order supersymmetric invariant when $N$ is treated as a composite field given by (\ref{CompositeLinear}). It is reasonable to expect that this can be done by a single linear multiplet action, which can be obtained by setting $\cF_{11} = L^{-1}$ in the action (\ref{LinearConformalAction}). There is, however, a problem with this choice; the action does not include a pure $N^2$ term, but includes $L^{-1} N^2$, thus not producing a desired kind of action, i.e. the supersymmetric completion of an $F^4$ term. One can evade this problem by considering two linear multiplets, one of which will be used to gauge fix the conformal symmetries and the other as the composite multiplet (\ref{CompositeLinear}). We let two multiplets be
\bea
(\underline{L}_{ij}, \underline{\vf}_i, \underline{E}^a, \underline{N}) \quad \text{and} \quad (L_{ij}, \vf_i, E^a, N)  \,,
\eea
where the first multiplet is the composite multiplet with the composite expressions being (\ref{CompositeLinear}), and the second one is the compensator. We, then, make our choice for $\cF_{AB}$ as
\bea
&& \cF_{11} = L^{-1} \,, \qquad \cF_{12} = - L^{-3} L_{ij} \underline{L}^{ij} \,, \qquad \cF_{21} = \cF_{22} = 0 \,,\nn\\
&& \cF_{121ij} = \cF_{112ij} = - L^{-3} L_{ij} \,, \qquad L_{122ij} = - L^{-3} \underline{L}_{ij} + 3 L^{-5} L_{ij} L_{kl} \underline{L}^{kl}
\eea
where  $``1"$ is used for the underlined composite linear multiplet and $``2"$ representing the compensating multiplet. With these choices, the bosonic part of the lowest component of the vector multiplet (\ref{VectorCompositeSUGRA}) is given by
\bea
\r &=& L^{-1} \underline{N} - L^{-3} L_{ij} \underline{L}^{ij} N \,.
\label{Compr}
\eea
Consequently, the density formulae
\bea
e^{-1} \cL &=& \r \underline{N} + Y_{ij} \underline{L}^{ij} + \rmi \bar{\l} \underline{\vf}  + A_a \underline{E}^a  \nn\\
&& - \ft12 \bar{\p}_\m^i \g^\m \l^i \underline{L}_{ij} + \ft12 \r \p_\m \g^\m \underline{\vf} + \ft14 \rmi \r \bar{\p}_\m^i \g^{\m\n} \p_\n^j \underline{L}_{ij} \,, 
\eea
would produce a linear multiplet action such that it would give rise to the supersymmetric completion of the $F^4$ term upon using the composite formulae (\ref{CompositeLinear}).  In order to produce the non-Abelian extension of the off-shell higher derivative vector multiplet action, one can follow the two derivative example \cite{Kugo:2000af}, which is based on a Noether procedure. Upon non-Abelian extension, the only field that picks up a $g-$dependent term in its supersymmetry transformation rule is $Y_{ij}$ \cite{Bergshoeff:2002qk}
\bea
Y_{ij}^I |_g &=& -\ft12 \rmi g \bar\e^{(i} f_{JK}{}^I \r^J \l^{j)K} \,,
\eea
where $I = 1,\ldots, n$ is the Yang-Mills index. The $g-$dependent terms from the variation of $Y_{ij}^I$ must be annihilated by the gauge covariantization of the superconformal derivatives $\widehat{F}_{\m\n}, \cD_\m \r, \cD_\m \l^i$ and $\cD_\m Y_{ij}$, and possible additional explicit $g-$dependent terms. Thus, keeping only the track of the $g-$terms, it is straightforward to find the required modification to the vector multiplet action. We will address this question, as well as the relation between the five dimensional and six dimensional higher derivative Yang-Mills multiplet invariants \cite{Butter:2016qkx} in a forthcoming paper.

%%%%%%%%%%%%%%%%%%%%%%%%%%%%%%%%%%%%
\section{Conclusions}{\label{Conclusions}}
%%%%%%%%%%%%%%%%%%%%%%%%%%%%%%%%%%%%

In this paper, we present a detailed study on the linear multiplets of the five dimensional $\cN=2$ supersymmetry. To construct the linear multiplet actions for $N-$number of interacting linear multiplets, we made extensive use of the superconformal tensor calulus. Our main result for the rigid case is presented in equation (\ref{SimplifiedRigid}), and for the conformal supergravity coupling, the main result is presented in the equation (\ref{LinearConformalAction}). As a byproduct of our formulation, we also describe the construction of off-diagonal supergravity invariants, and presented two particular examples: the $RN$ and the $RN^3$ invariants.

There are number of directions to pursue following our work. First of all, the formulation we present here can be directly generalized to six dimensional $\cN = (1,0)$ conformal supergravity. As the vector multiplet of the six dimensional supergravity does not include a scalar field as we have in here, we expect that the function $\cF_{AB}$ that determines the interaction of the linear multiplet is less constrained. Furthermore, the construction procedure of higher derivative vector multiplet action that we described here can be generalized to six dimensions, however this time it generates the supersymmetric completion of the $F \Box F$ invariant, which, upon non-Abelian generalization, would produce the conformal Yang-Mills supersymmetry in six dimensions. The vacuum solutions and the spectrum corresponding to the off-diagonal extended (higher derivative) supergravity also remains to be investigated. For the construction of  five dimensional ghost-free higher derivative gauged supergravity model, the off-diagonal invariants given in this paper are expected to play an important role in eliminating the ghost degrees of freedom. Finally, it is reasonable to expect that the off-diagonal invariants constructed in this paper can play a role in higher order corrections to black hole entropy or higher order effects in AdS/CFT, see i.e. \cite{deWit:2009de,Cremonini:2009sy}. On the other hand, one can also argue that as the auxiliary fields can be eliminated perturbatively even when they acquire kinetric terms via higher derivative invariants, the off-diagonal invariants may completely be absorbed into the on-shell two derivative theory by means of a perturbative elimination \cite{Argyres:2003tg} and field redefinitions \cite{Myers:2009ij}. Therefore, it would be interesting to see explicitly whether off-diagonal invariants can survive such an elimination procedure and lead to non-trivial higher order effects in AdS/CFT.

%%%%%%%%%%%%%%%%%%%%%%%%%%%%%%%%%%%%
\section*{Acknowledgement}
%%%%%%%%%%%%%%%%%%%%%%%%%%%%%%%%%%%%
It is a pleasure to thank Eric Bergshoeff, Yi Pang and Ergin Sezgin for useful discussions. The work of MO is supported in part by Marie Curie Cofund (No.116C028).
%%%%%%%%%%%%%%%%%%%%%%%%%%%%%%%%%%%%

\appendix

%%%%%%%%%%%%%%%%%%%%%%%%%%%%%%%%%%%%
\section{Weyl Multiplets of Five Dimensional $\cN=2$ Supergravity}{\label{Appendix1}}
%%%%%%%%%%%%%%%%%%%%%%%%%%%%%%%%%%%%

The five dimensional $\cN=2$ conformal tensor calculus is based on the exceptional superalgebra $F^2 (4)$ with the generators
\bea
P_a \,, \quad M_{ab} \,, \quad D \,, \quad K_a \,, \quad U_{ij}  \,, \quad Q_{\a i} \,, \quad S_{\a i} \,,
\eea
with the corresponding gauge fields
\bea
e_\m{}^a \,, \quad \o_\m{}^{ab} \,, \quad b_\m \,, \quad f_\m{}^a \,, \quad V_\m^{ij} \,, \quad  \p_\m^i \,, \quad  \f_\m^i \,,
\eea
where $a,b, \ldots$ are the Lorentz indices $\m,\n,\ldots$ are the world vector indices, $\a$ is a spinor index and $i =1,2$ is an $\SU(2)$ index. Here, $\{P_a, M_{ab}, D, K_a\}$ are the generators of the usual conformal algebra, whereas $U_{ij}$ is the $SU(2)$ generator and $Q_{\a i}$ and $S_{\a i}$ are the generators of the $Q-$SUSY and the $S-$SUSY respectively. In order to relate the $P_a$ and $M_{ab}$ to the diffeomorphisms of spacetime, one can impose the so-called conventional curvature constraints \cite{Bergshoeff:2001hc}, in which case the the fields $\o_\m{}^{ab}, f_\m^a$ and $\f_\m^i$ becomes dependant fields, and by a simple degree of freedom counting shows that there are $21$ bosonic and $24$ fermionic degrees of freedom. In order to match the bosonic and fermionic degrees of freedom, and to have an off-shell closed superconformal multiplet, one can add two set of distinct matter fields. One of the choices, a real scalar $D$, an anti-symmetric tensor $T_{ab}$ and a symplectic Marojana spinor $\chi^i$ would lead to the standard Weyl multiplet.  The $Q$-, $S$- and $K$- transformations of these fields are given by
\bea
\d e_\m{}^a   &=&  \ft 12\bar\e \g^a \psi_\m  \nn\, ,\\
\d \psi_\m^i   &=& (\partial_\m+\tfrac{1}{2}b_\m+\tfrac{1}{4}\omega_\m{}^{ab}\g_{ab})\e^i-V_\m^{ij}\e_j + \rmi \g\cdot T \g_\m
\e^i - \rmi \g_\m
\eta^i  \nn\, ,\\
\d V_\m{}^{ij} &=&  -\ft32\rmi \bar\e^{(i} \phi_\m^{j)} +4
\bar\e^{(i}\g_\m \chi^{j)}
+ \rmi \bar\e^{(i} \g\cdot T \psi_\m^{j)} + \ft32\rmi
\bar\eta^{(i}\psi_\m^{j)} \nn\, ,\\
\d T_{ab} &=& \tfrac12 \rmi\bar\e \g_{ab} \chi - \tfrac3{32} \rmi \bar\e \widehat{R}_{ab}(Q)\,, \nn\\
\d \chi^i &=& \tfrac14 \e^i D - \tfrac1{64} \g \cdot  \widehat{R}^{ij}(V) \e_j + \tfrac18 \rmi \g^{ab}\slashed{\mathcal{D}}T_{ab}\e^i - \tfrac18 \rmi \g^a  \mathcal{D}^b T_{ab} \e^i \nn\\
&& - \tfrac14 \g^{abcd}T_{ab} T_{cd} \e^i + \tfrac16 T^2 \e^i + \tfrac14 \g \cdot T \eta^i\,, \nn\\
\d D &=& \bar\e \slashed{\mathcal{D}}\chi - \tfrac53 \rmi \bar\e \g \cdot T \chi - \rmi \bar\eta \chi\,, \nn\\
\d b_\m       &=& \ft12 \rmi \bar\e\phi_\m -2 \bar\e\g_\mu \chi +
\ft12\rmi \bar\eta\psi_\mu+2\Lambda _{K\mu } \,,
\label{SWMTR}
\eea
where
\bea
\mathcal{D}_\m\chi^i&=&(\partial_\mu - \tfrac32 b_\mu +\tfrac14 \omega_\mu{}^{ab} \g_{ab})\chi^i -V_\mu^{ij}\chi_j
- \tfrac14 \p_\m^i D  + \tfrac1{64} \g \cdot  \widehat{R}^{ij}(V) \p_{\m j}\nn\\
&& - \tfrac18 \rmi \g^{ab}\slashed{\mathcal{D}}T_{ab}\p_\m^i + \tfrac18 \rmi \g^a  \mathcal{D}^b T_{ab} \p_\m^i  + \tfrac14 \g^{abcd}T_{ab} T_{cd} \p_\m^i - \tfrac16 T^2 \p_\m^i - \tfrac14 \g \cdot T \phi_\m^i\,, \nn \\
\mathcal{D}_\m T_{ab}&=&\partial_\m T_{ab}-b_\m T_{ab}-2\omega_\mu{}^c{}_{[a}T_{b]c}-\tfrac{1}{2}i\bar{\p}_\m\g_{ab}\chi+\tfrac{3}{32}i\bar{\p}_\m\widehat{R}_{ab}(Q)\,.
\eea
The supercovariant curvatures that are relevant to our discussion here are given by \cite{Bergshoeff:2001hc} 
\bea
R_{\m\n}{}^a (P) &=& 2 \partial_{[\m} e_{\n]}{}^a  + 2 \o_{[\m}{}^{ab}  e_{\n]b} + 2 b_{[\m} e_{\n]}{}^a - \ft12 \bar{\p}_{[\m} \g^a \p_{\n]} \,,\nn\\ 
R_{\m\n}(D) &=& 2 \partial_{[\m} b_{\n]} - 4 f_{[\m}^a e_{\n]a} - \rmi \bar{\f}_{[\m} \p_{\n]} \,,\nn\\ 
\widehat{R}_{\m\n}{}^{ab}(M)&=&2\partial_{[\m}\o_{\n ]}{}^{ab}+2\o_{[\m}{}^{ac}\o_{\n ]c}{}^{b} + 8 f_{[\m}{}^{[a}e_{\n ]}{}^{b]}+\rmi \bar\p_{[\m}\g^{ab}\p_{\n ]} + \rmi \bar\p_{[\m}\g^{[a} \g \cdot T \g^{b]}\p_{\n ]}  \nn\\
&& +\bar\p_{[\m} \g^{[a} \widehat{R}_{\n ]}{}^{b]}(Q)+\tfrac12 \bar\p_{[\m}\g_{\n ]} \widehat{R}^{ab}(Q) -8 \bar\p_{[\m} e_{\n ]}{}^{[a} \g^{b]}\chi+i\bar{\phi}_{[\m} \g^{ab} \p_{\n]}\nn\,, \\
\widehat{R}_{\m\n}{}^{ij}(V)&=&2\partial_{[\m} V_{\n]}{}^{ij} -2V_{[\m}{}^{k( i}
V_{\n ]\,k}{}^{j)} {-3\rmi}{\bar\f}^{( i}_{[\m}\p^{j)}_{\n ]}  - 8 \bar{\p}^{(i}_{[\m}
\g_{\n]} \chi^{j)} - \rmi \bar{\p}^{(i}_{[\m} \g\cdot T \psi_{\n]}^{j)}   \,, \\
\widehat{R}_{\m\n}^i(Q)&=&2\partial_{[\m}\p_{\n]}^i+\frac{1}{2}\o_{[\m}{}^{ab}\g_{ab}\p_{\n]}^i+b_{[\m}\p_{\n]}^i-2V_{[\m}^{ij}\p_{\n] j}-2i\g_{[\m}\phi_{\n]}^i+2i\g\cdot T\g_{[\m}\p_{\n]}^i\,, \nn
\eea
where the spin connection $\omega_{\m}{}^{ab}$, the $S$-supersymmetry gauge field $\phi_\m^i$ and the special conformal symmetry gauge field $f_\m{}^a$ are composite fields. Their explicit expressions are given by
\bea \o_\m{}^{ab}
&=& 2 e^{\n[a} \partial_{[\m} e_{\n]}^{~b]} - e^{\n[a} e^{b]\s} e_{\m c}
\partial_\n e^{~c}_\s
+ 2 e_\m^{~~[a} b^{b]} - \ft12 \bar{\p}^{[b} \g^{a]} \p_\m - \ft14
\bar{\p}^b \g_\m \p^a \,,\nonumber\\
\f^i_\m &=& \ft13\rmi \g^a \widehat{R}^\prime _{\m a}{}^i(Q) - \ft1{24}\rmi
\g_\m \g^{ab} \widehat{R}^\prime _{ab}{}^i(Q)\,, \label{transfDepF} \\
f^a_\m &=&  - \ft16{\cal R}_\mu {}^a + \ft1{48}e_\mu {}^a {\cal R},\quad {\cal R}_{\mu \nu }\equiv \widehat{R}_{\mu \rho }^{\prime~~ab}(M) e_b{}^\rho
e_{\nu a},\quad {\cal R}\equiv {\cal R}_\mu {}^\mu\,, \nonumber
\label{cf1}
\eea
where the notation $\widehat{R}_{ab}'(Q)$ and $\widehat{R}_{\mu \rho }^{\prime~~ab}(M)$ implies that these expressions are obtained from $\widehat{R}_{ab}(Q)$ and $\widehat{R}_{\mu \rho }{}^{~~ab}(M) $ by omitting the $\phi_\m^i$ and $f_{\m}^a$ terms respectively. These composite expressions can be obtained by the following conventional constraints
\bea
R_{\m\n}{}^a (P) &=& 0 \,,\nn\\ 
e^\m{}_a \widehat{R}_{\m\n}{}^{ab}(M) &=& 0 \nn\\
\g^\m \widehat{R}_{\m\n}^i (Q) &=& 0 \,.
\eea
Note that these constraints imply relations between different curvatures by Bianchi identities, i.e.
\bea
e_{[\m}{}^a {R}_{\n\r]}(D) &=& \widehat{R}_{[\m\n\r]}{}^a (M) \,, \qquad {R}_{\m\n} (D) = 0 \,.
\eea

Instead of working with the matter sector of the standard Weyl multiplet, one can add the gauge sector of the Weyl multiplet a vector field $C_\m$, a two-form gauge field $B_{\m\n}$, a dilaton field $\s$ and a dilatino $\p^i$. This would form the dilaton Weyl multiplet. The $Q$-, $S$- and $K$- transformation rules of the fields in the dilaton Weyl multiplet are given by \cite{Bergshoeff:2001hc}
\bea
\d e_\m{}^a   &=&  \ft 12\bar\e \g^a \psi_\m  \nn\, ,\\
\d \psi_\m^i   &=& (\partial_\m+\tfrac{1}{2}b_\m+\tfrac{1}{4}\omega_\m{}^{ab}\g_{ab})\e^i-V_\m^{ij}\e_j + \rmi \g\cdot \underline{T} \g_\m
\e^i - \rmi \g_\m
\eta^i  \nn\, ,\\
\d V_\m{}^{ij} &=&  -\ft32\rmi \bar\e^{(i} \phi_\m^{j)} +4
\bar\e^{(i}\g_\m \underline{\chi}^{j)}
+ \rmi \bar\e^{(i} \g\cdot \underline{T} \psi_\m^{j)} + \ft32\rmi
\bar\eta^{(i}\psi_\m^{j)} \,, \nn\\
\d C_\m
&=& -\ft12\rmi \s \bar{\e} \p_\m + \ft12
\bar{\e} \g_\m \p, \nn\\
\d B_{\m\n}
&=& \ft12 \s^2 \bar{\e} \g_{[\m} \p_{\n]} + \ft12 \rmi \s \bar{\e}
\g_{\m\n} \p + C_{[\m} \d(\e) C_{\n]}, \nonumber\\
\d \p^i &=& - \ft14 \g \cdot \widehat{G} \e^i -\ft12\rmi \slashed{\mathcal{D}} \s
\e^i + \s \g \cdot \underline{T} \e^i -\ft14\rmi\s^{-1}\e_j \bar\p^i \p^j  + \s\eta^i \,,\nonumber\\
\d \s &=& \ft12 \rmi \bar{\e} \p \, ,\nonumber\\
\d b_\m       &=& \ft12 \rmi \bar\e\phi_\m -2 \bar\e\g_\mu \underline{\chi} +
\ft12\rmi \bar\eta\psi_\mu+2\Lambda _{K\mu } \,,
\label{TransDW}
\eea
where
\begin{eqnarray}
\mathcal{D}_\mu\, \s &=&  (\partial_\mu - b_\mu) \s
- \tfrac12\, \rmi\bar{\psi}_\mu \p \ ,
\nn\\
\mathcal{D}_\mu \p^i &=&  (\partial_\mu -\ft32 b_\mu +\ft14\,  \o_\mu{}^{ab}\g_{ab} ) \p^{i} - V_\mu^{ij} \p_j +\tfrac 14 \g
\cdot \widehat{G} \p_\mu^i  \nn\\
&& + \ft12\rmi \slashed{\cD} \s \p_\mu^i
+\ft14\rmi\s^{-1}\p_{\m j}\bar\p^i\p^j  - \s  \g \cdot \underline{T} \p_\mu^i - \s
\phi_\mu^i\,,
\label{cd1}
\end{eqnarray}
and the supercovariant curvatures are defined according to
\bea
\widehat{G}_{\m\n}  &=& 2 \partial_{[\mu } C_{\nu ]} - \bar{\p}_{[\m} \g_{\n]} \p + \tfrac 12 \rmi
\s \bar{\p}_{[\m} \p_{\n]} \label{hatG} ,\nn\\
\widehat{H}_{\m\n\r} &=&  3\partial _{[\mu }B_{\nu \rho ]} + \ft32 C_{[\m}  G_{\n\r]} - \ft34
\s^2 \bar{\p}_{[\m} \g_\n \p_{\r]} - \ft32\rmi \s \bar{\p}_{[\m}
\g_{\n\r]} \p.
\label{DefH}
\eea
The underlined expressions $\underline{T}^{ab}, \underline{\chi}^i$ and $\underline{D}$, are the matter fields of the standard Weyl multiplet. However, in the transformation rules above, they are expressed in terms of the fields of the dilaton Weyl multiplet as \cite{Bergshoeff:2001hc}
\bea
\underline{T}^{ab} &=& \ft18 \s^{-2} \Big( \s \widehat G^{ab} + \ft16 \e^{abcde} \widehat H_{cde} + \ft14 \rmi \bar\p \g^{ab} \p \Big) \,, \nn\\
\underline{\chi}^i &=& \ft18 \rmi \s^{-1} \slashed{\cD} \p^i + \ft1{16} \rmi \s^{-2} \slashed{\cD} \s \p^i - \ft1{32} \s^{-2} \g \cdot \widehat G \p^i + \ft14 \s^{-1} \g \cdot \underline{T} \p^i \nn\\
&& + \ft1{32} \rmi \s^{-3} \p_j \bar\p^i\p^j ,\,\nn\\
\underline{D} &=& \ft14 \s^{-1} \Box^c \s + \ft18 \s^{-2} (\cD_a \s) (\cD^a \s) - \ft1{16} \s^{-2} \widehat G_{\m\n} \widehat G^{\m\n}\nn\\
&& - \ft18 \s^{-2} \bar\p \slashed{\cD} \p  -\ft1{64} \s^{-4} \bar\p^i \p^j \bar\p_i \p_j - 4 \rmi \s^{-1} \p \underline{\chi} \nn\\
&& + \Big( - \ft{26}3 \underline{T_{ab}} + 2 \s^{-1} \widehat G_{ab} + \ft14 \rmi \s^{-2} \bar\p \g_{ab} \p \Big) \underline{T}^{ab}\,,
\label{UMap}
\eea
where the superconformal d'Alambertian for $\s$ is given by
\bea
&&\Box^c \s= (\partial^a - 2b^a + \o_b{}^{ba}) \cD_a \s - \ft12 \rmi \bar\p_a \cD^a\psi - 2\sigma \bar\p_a \g^a {\chi} \nn\\
&&\quad\quad + \ft12 \bar\p_a \g^a \g \cdot \underline{T} \psi + \ft12 \bar\phi_a \g^a \psi + 2 f_a{}^a \sigma \,.
\eea

\providecommand{\href}[2]{#2}\begingroup\raggedright\endgroup

\end{document}